\documentclass[final]{aipproc}
\usepackage{amssymb}
\usepackage{amsmath}
\layoutstyle{6x9}

\newcommand{\beeq}{\begin{equation}}
\newcommand{\eneq}{\end{equation}}
\newcommand{\beeqa}{\begin{eqnarray}}
\newcommand{\eneqa}{\end{eqnarray}}

\begin{document}

\begin{flushright}
IFUP-TH 2004/39, UPRF-2004-29 \\
\end{flushright}

\title{Study of dynamical supersymmetry breaking for the two dimensional lattice Wess-Zumino model
\protect\footnote{Talk presented by A. Feo.}
}

\author{Matteo Beccaria}{
  address={INFN, Sezione di Lecce, and
Dipartimento di Fisica dell'Universit\`a di Lecce,
Via Arnesano, ex Collegio Fiorini, I-73100 Lecce, Italy}
}

\author{Gian Fabrizio De Angelis}{
  address={INFN, Sezione di Lecce, and
Dipartimento di Fisica dell'Universit\`a di Lecce,
Via Arnesano, ex Collegio Fiorini, I-73100 Lecce, Italy}
}

\author{Massimo Campostrini}{
  address={INFN, Sezione di Pisa, and
Dipartimento di Fisica ``Enrico Fermi'' dell'Universit\`a di Pisa,
Via Buonarroti 2, I-56125 Pisa, Italy}
}

\author{Alessandra Feo}{
  address={Dipartimento di Fisica, Universit\`a di Parma and INFN Gruppo Collegato di Parma,
Parco Area delle Scienze, 7/A, 43100 Parma, Italy}
}

\begin{abstract}
A new approach to the study of the transition point in a class
of two dimensional Wess-Zumino models is presented. The method is based on
the calculation of rigorous lower bounds on the ground state energy density
in the infinite lattice limit. Such bounds are useful in the discussion of
supersymmetry phase transition. The transition point is then determined
and compared with recent results based on large-scale Green Function Monte
Carlo simulations with good agreement.
\end{abstract}

\maketitle

The simplest theoretical laboratory to study non perturbative dynamical supersymmetry breaking
is the two dimensional $N=1$ Wess-Zumino model that involves chiral superfields with no vector
multiplets. Let us remind the (continuum) $N=1$ supersymmetry algebra, 
$\{Q_\alpha,Q_\beta\} = 2(\not{P} C)_{\alpha\beta}$. Since $P_i$ are not conserved on the lattice,
a lattice formulation of a supersymmetric model must break this algebra explicitly.
A very important advantage of the Hamiltonian formulation is the possibility to conserve  
exactly a key subalgebra of this relation; specializing to $1+1$ dimensions, in a Majorana
basis, $\gamma_0 = C = \sigma_2$, $\gamma_1 = i \sigma_3$, the algebra becomes 
$Q_1^2 = Q_2^2 = P^0 \equiv H, \qquad \{Q_1,Q_2\} = 2 P^1 \equiv 2 P $. 
On the lattice, since $H$ is conserved but $P$ is not, we can pick up 
one of the supercharges, say, $Q_1^2 = H$, build a discretized version $Q_L$ and define 
the lattice Hamiltonian $H = Q_L^2$.

The explicit lattice model is built by considering a spatial 
lattice with $L$ sites. On each site we place 
a real scalar field $\varphi_n$ together with its conjugate momentum $p_n$ such 
that $[p_n, \varphi_m] = -i\delta_{n,m}$.  The associated fermion is a Majorana fermion
$\psi_{a, n}$ with $a=1,2$ and 
$\{\psi_{a, n}, \psi_{b, m}\} = \delta_{a,b}\delta_{n,m}$, 
$\psi_{a,n}^\dagger = \psi_{a,n}$. 
The discretized supercharge 
$Q_L = \sum_{n=1}^L [p_n\psi_{1,n}- (\frac{\varphi_{n+1}-\varphi_{n-1}}{2}
+ V(\varphi_n) )\psi_{2,n} ] $, 
with arbitrary $V(\varphi)$ (called prepotential) can be used to define a 
semipositive definite lattice Hamiltonian $H = Q_L^2$. 
Notice that $Q_1^2 = H$ is enough to guarantee that 
$E_0 \equiv \langle \Psi_0 | H | \Psi_0 \rangle \ge 0$, that all eigenstates of
$H$ with $ E > 0$ are paired in doublets and that $E_0=0$ if and only if supersymmetry 
is unbroken, i.e., the ground state is anihilated by $Q_1$.

Rigorous results from the continuum can be found in \cite{jaffe}. 
On the lattice, accurate numerical results are available~\cite{wz1,beccaria}, although
a clean determination of the supersymmetry breaking transition remains rather elusive.
All predictions and results indicate: for the model with cubic prepotential, 
$V=\varphi^3$, unbroken supersymmetry; for the model with a quadratic prepotential, 
$V=\lambda_2\varphi^2 + \lambda_0 $, dynamical supersymmetry breaking. Along 
a line of constant $\lambda_2$ the results for a quadratic potential indicate the existence of two phases:
a phase of broken supersymmetry with unbroken discrete $Z_2$ at high $\lambda_0$ and 
a phase of unbroken supersymmetry with broken $Z_2$ at low $\lambda_0$,  
separated by a single phase transition.
On the other hand, strong coupling expansion demonstrate that for a polynomial $V(\varphi)$, 
the relevant parameter is just its degree $q$.
For odd $q$, strong coupling expansion and tree-level results agree and supersymmetry is 
expected to be unbroken.
This conclusion gains further support from the nonvanishing value of the Witten index \cite{witten}.
For even $q$ in strong coupling expansion, the ground state has a positive energy density also for 
$L \to \infty $ and supersymmetry appears to be broken for all $\lambda_0$.
On the other hand, weak coupling expansion predicts unbroken supersymmetry when $\lambda_0 < 0$ \cite{wz1}.

We used two different approaches to investigate the pattern of dynamical supersymmetry breaking
in a class of two dimensional Wess-Zumino models.
In the first one, \cite{wz1,beccaria}, the numerical simulations were performed using the 
Green Function Monte Carlo (GFMC) algorithm and strong coupling expansion.
The GFMC is a method that computes a numerical representation of the ground state energy density 
on a finite lattice with $L$ sites in terms of the states carried by an ensemble of $K$ walkers.
By performing numerical simulations along a line of constant $\lambda_2=0.5$,
we determined the numerical value of $\lambda_0$ separating a phase of broken 
supersymmetry from a phase of unbroken supersymmetry.
The usual technique for the study of a phase transition is the crossing method applied to the 
Binder cumulant \cite{wz1}. The crossing method consists in plotting $B$ vs.\ $\lambda_0$ for
several values of $L$. The crossing point $\lambda_0^{\rm cr}(L_1,L_2)$, determined by the condition 
$B(\lambda_0^{\rm cr},L_1) = B(\lambda_0^{\rm cr},L_2)$ is an estimator of $\lambda_0^{(c)}$.
The value obtained is $\lambda_0^{(c)}=-0.48\pm0.01$.
The main source of systematic errors in this method is the need to extrapolate to infinity
both $K$ and $L$. For this reason, an indepentend method to test the numerical results of 
\cite{wz1} is welcome.

The second method is based on a new approach to the study of the supersymmetry phase diagram 
introduced in Ref. \cite{wz2} and is based on the calculation of rigorous lower bounds 
on the ground state energy density in the infinite lattice limit.  
The Hamiltonian lattice version of the Wess-Zumino model conserves enough supersymmetry 
to prove that the ground state has a non negative energy density $\rho\ge 0$, as its continuum limit.
Moreover, the ground state is supersymmetric if and only if $\rho=0$,
whereas it breaks (dynamically) supersymmetry if $\rho>0$. 
Therefore, if an exact positive lower bound $\rho_{\rm LB}$ is found with 
$0< \rho_{\rm LB} \le \rho$, we can claim that supersymmetry is broken.

The relevant quantity for our analysis is the ground state energy density
$\rho$ evaluated on the infinite lattice limit,  
$\rho = \lim_{L\to\infty}\frac{E_0(L)}{L} = \lim_{L\to\infty}\rho(L)$.
It can be used to tell between the two phases of the model: supersymmetric with $\rho=0$ or 
broken with $\rho > 0$.
We have shown in \cite{wz2} how to build a sequence of bounds $\rho^{(L)}$
which are the ground state energy density of the Hamiltonian H with modified couplings on a cluster of $L$ sites.
We now explain how to exploit the sequence of bounds $\rho^{(L)}$ 
to determine the phase at a particular point in the coupling constant space. In order to do so, 
we compute numerically $\rho^{(L)}$ at various values of the cluster size $L$. 
If $\rho^{(L)}>0$ for some $L$,
we can immediately conclude that we are in the broken phase. 
On the other hand, if we find a negative 
bound we cannot conclude in which phase we are. However, we know that $\rho^{(L)}\to \rho$ for $L\to\infty$
and the study of $\rho^{(L)}$ as a function of both $L$ and the coupling constants permit the identification 
of the phase in all cases. To test the method we studied in detail the case of the quadratic prepotential
and discussed the dependence of $\rho$ on $\lambda_0$ at a fixed value of $\lambda_2 = 0.5$ (as in Ref. \cite{wz1}). 

\begin{figure}
\includegraphics[width=.3\textheight]{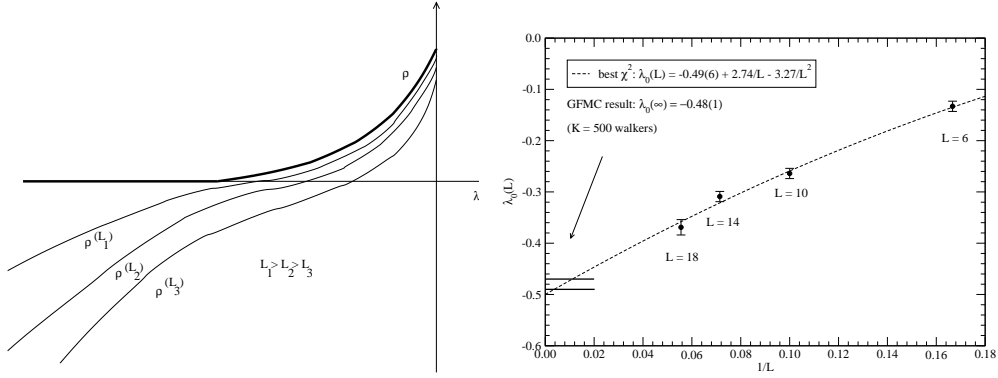}\includegraphics[width=.3\textheight]{Plot.extrapolation2.eps}
\label{fig}
\caption{Left: Qualitative plot of the function $\rho(\lambda_0)$ and $\rho^{(L)}(\lambda_0)$. 
Right: Best fit with a quadratic polynomial in $1/L$ together with the best GFMC result \cite{wz1} 
obtained with $K=500$ walkers.}
\end{figure}

On the left side of Fig.~\ref{fig} a reasonable qualitative pattern of the curves representing 
$\rho^{(L)}(\lambda_0)$ is shown. Notice that a single zero is expected in $\rho^{(L)}(\lambda_0)$
at some $\lambda_0 = \lambda_0(L)$. Since $\lim_{L\to\infty}\rho^{(L)} = \rho$, we expect that 
$\lambda_0(L)\to\lambda_0^*$ for $L\to \infty$, 
allowing for a determination of the critical coupling $\lambda_0^*$.
The results of the energy for the lower bound $\rho^{(L)}(\lambda_0)$ for all cluster sizes
behaves as expected: it is positive around $\lambda_0 = 0$ and 
decreases as $\lambda_0$ moves to the left. At a certain unique point 
$\lambda_0^*(L)$, the bound vanishes and 
remains negative for $\lambda_0 < \lambda_0^*(L)$. This means that supersymmetry breaking can be excluded 
for $\lambda_0 > \min_L\lambda_0^*(L)$. Also, consistency of the bound means that $\lambda_0^*(L)$ must converge
to the infinite-volume critical point as $L\to\infty$. 
Since the difference between the exact Hamiltonian and the 
one used to derive the bound is ${\cal O}(1/L)$, we can fit $\lambda_0^*(L)$ with a polynomial in $1/L$. 
This is shown on the right side of Fig.~\ref{fig} \cite{wz2}, where also the GFMC result is quoted. 
The best fit with a parabolic function gives $\lambda_0^* = -0.49\pm 0.06$, quite in agreement with 
the previous $\lambda_{0, \rm GFMC}^* = -0.48\pm0.01$ using the GFMC algorithm.

\end{document}